\documentclass[12pt]{article}
\usepackage[utf8]{inputenc}
\usepackage{amsmath,amsthm,amssymb}
\usepackage{tabularx}
\usepackage{bbding}
\usepackage{graphicx}
\graphicspath{ {./images/} }
\usepackage{indentfirst}
\usepackage[document]{ragged2e}
\usepackage[none]{hyphenat}
\usepackage{bm,multirow,float}
\usepackage{authblk,url}
\usepackage[a4paper, margin=1in]{geometry}
\usepackage{multirow}
\usepackage{verbatim}
\immediate\write18{texcount -tex -sum  \jobname.tex > \jobname.wordcount.tex}
\usepackage{color, soul}
\usepackage{longtable}
\usepackage{booktabs}
\usepackage{setspace}
\usepackage{array}

\usepackage[colorinlistoftodos]{todonotes}
\newcolumntype{L}[1]{>{\raggedright\let\newline\\\arraybackslash\hspace{0pt}}m{#1}}
\newcolumntype{C}[1]{>{\centering\let\newline\\\arraybackslash\hspace{0pt}}m{#1}}
\newcolumntype{R}[1]{>{\raggedleft\let\newline\\\arraybackslash\hspace{0pt}}m{#1}}

\providecommand{\keywords}[1]
{
  \small	
  \textbf{\textit{Keywords---}} #1
}

\title{BayCANN: Streamlining Bayesian Calibration with Artificial Neural Network Metamodeling}
\author[1,*]{Hawre Jalal} 
\author[2,*]{Fernando Alarid-Escudero}

\affil[1]{Department of Health Policy and Management, University of Pittsburgh, Graduate School of Public Health,Pittsburgh, PA, 15261}
\affil[2]{Division of Public Administration, Center for Research and Teaching in Economics (CIDE), Aguascalientes, Aguascalientes, Mexico, 20313}

\date{}

\usepackage[parfill]{parskip} 

\begin{document}

\maketitle

\begin{abstract}
Purpose: Bayesian calibration is theoretically superior to standard direct-search algorithm because it can reveal the full joint posterior distribution of the calibrated parameters.  However, to date, Bayesian calibration has not been used often in health decision sciences due to practical and computational burdens.  In this paper we propose to use artificial neural networks (ANN) as one solution to these limitations.   

Methods: Bayesian Calibration using Artificial Neural Networks (BayCANN) involves (1) training an ANN metamodel on a sample of model inputs and outputs, and (2) then calibrating the trained ANN metamodel instead of the full model in a probabilistic programming language to obtain the posterior joint distribution of the calibrated parameters. We demonstrate BayCANN by calibrating a natural history model of colorectal cancer to adenoma prevalence and cancer incidence data. In addition, we compare the efficiency and accuracy of BayCANN against performing a Bayesian calibration directly on the simulation model using an incremental mixture importance sampling (IMIS) algorithm. 

Results: BayCANN was generally more accurate than IMIS in recovering the "true" parameter values. The ratio of the absolute ANN deviation from the truth compared to IMIS for eight out of the nine calibrated parameters were less than one indicating that BayCANN was more accurate than IMIS.  In addition, BayCANN took about 15 minutes total compared to the IMIS method which took 80 minutes.

Conclusions: In our case study, BayCANN was more accurate than IMIS and was five-folds faster.  Because BayCANN does not depend on the structure of the simulation model, it can be adapted to models of various levels of complexity with minor changes to its structure.  We provide BayCANN's open-source implementation in R.

\end{abstract}


* Please address correspondences to either Hawre Jalal (\texttt{hjalal@pitt.edu}) or Fernando Alarid-Escudero (\texttt{fernando.alarid@cide.edu}).

\section{Background}
Modelers and decision-makers often use mathematical or simulation models to simplify real-life complexity and inform decisions, particularly those for which uncertainty is inherent.  
However, some of the model parameters might be either unobserved or unobservable due to various financial, practical or ethical reasons. For example, a model that simulates the natural history of cancer progression may lack an estimate for the rate at which an individual transitions from a pre-symptomatic cancer state to becoming symptomatic.  Although this rate might not be directly observable, mathematical models have proven useful in reconstructing cancer progression dynamics and estimating these unknown parameters using a technique commonly referred to as calibration. Calibration involves modifying the model input parameters until a desired output is obtained. Thus, researchers have successfully calibrated their models to other rich data on cancer such as prevalence of precancerous lesions or cancer incidence that are commonly produced from cancer models.\cite{Alarid-Escudero2018c, Vanni2011, Rutter2009}

Calibration has the potential of improving model structure, and recent guidelines recommend that model calibration should be performed where there is existing data on outputs.\cite{Weinstein2003a, Briggs2012} In addition, modelers are encouraged to report the uncertainty around calibrated parameters and use these uncertainties in both deterministic and probabilistic sensitivity analyses.\cite{Briggs2012} 

There are several calibration techniques with various levels of complexity.  For example, Nelder-Mead is a direct-search algorithms commonly used to calibrate models in health and medicine. Nelder-Mead is a deterministic approach that searches the parameter space for good-fitting parameter values.\cite{Nelder1965} Although Nelder-Mead is generally reliable, it cannot produce parameter distributions or directly inform correlations among the calibrated parameters.  It is also not guaranteed to produce a global optimal value because it might converge to a local optima.  
Unlike the direct-search algorithms, Bayesian methods are naturally suited for calibration because they can reveal the posterior joint and marginal distributions of the input parameters.\cite{Menzies2017} However, Bayesian methods are rarely implemented in practice due to the technical and computational challenges.  Bayesian calibration often requires tens or hundreds of thousands of simulations runs, and often requires a model written in a probabilistic programming language, such as Stan\cite{carpenter2017} or Bayesian inference Using Gibbs Sampling (BUGS)\cite{lunn2009bugs}.  We argue that the complexity of these tasks and their potential computational demand have prohibited a wider adoption of Bayesian calibration methods. 

In this manuscript, we propose using metamodels to streamline Bayesian calibration.  A metamodel is a second model that can be used to approximate the model's input-output relationship.\cite{Jalal2013} We specifically propose to use deep artificial neural networks (ANN) given their ability to map highly nonlinear relationship between sets of inputs and outputs. ANNs have been used as metamodels of both stochastic and deterministic responses mainly for their computational efficiency.\cite{Barton2009b,Badiru1998,R.D.Hurrion1997,Chambers2002,Zobel2008} One of the first implementation of ANN as metamodels was in 1992 for a scheduling simulation model.\cite{Pierreval1992,Pierreval1992a} Since then, ANNs have been successfully implemented as emulators of all sorts of discrete-event and continuous simulation models in a wide variety of fields.\cite{Kilmer1996,Sabuncuoglu2002,Fonseca2003,ElTabach2007} 
ANNs have also been proposed as proxies for nonlinear and simulation models\cite{Paiva2010,Mares2012,Pichler2003}. An example of ANNs as metamodels is estimating the mean and variance of patient time in emergency department visits.\cite{Kilmer1994,Kilmer1997}  Nowadays ANNs are widely popular as machine learning tools in artificial intelligence.\cite{schmidhuber2015deep} Deep learning using ANNs are used for visual recognition in self-driving cars\cite{ndikumana2020deep} and in classifying galaxies\cite{folkes1996artificial}.

ANNs have been used for calibration of computationally expensive models, such as general circulation and rainfall-runoff models in climate science\cite{Khu2004,Hauser2012}, and other complex global optimization techniques such as genetic algorithms.\cite{Wang2005}  Most algorithms use ANN mainly for its computational efficiency relative on the simulation models upon which they are built.  In this paper, we illustrate that in addition to its computational efficiency, ANN can also overcome practical challenges involved in representing the simulation model in probabilistic programming languages. We refer to our approach as Bayesian calibration via artificial neural networks, or BayCANN for short.

We demonstrate BayCANN by calibrating a realistic model of the natural history of colorectal cancer (CRC). We compare the results of this  approach to a full Bayesian calibration of the original model using an incremental mixture importance sampling (IMIS) algorithm. 
We provide the code in R and Stan for our application that researchers can use to calibrate their own models.

\section{Methods}
We start this exposition with reviewing Bayes formula and showing that Bayes formula can be applied directly to calibrate a model. We describe the computational burden of attempting to apply Bayes formula in most realistic models, and how deep ANNs can be used to streamline Bayesian calibration methods to calibrate these models. We illustrate this approach by calibrating a natural history model of CRC.  We also compare the performance of the BayCANN approach to a traditional Bayesian calibration using IMIS.

\subsection{Bayesian calibration}
In its most basic form, Bayes formula states 
\begin{equation}\label{eq:bayes}
	p(\theta|\text{data}) = \frac{l(\text{data}|\theta) p(\theta)}{p(data)}
\end{equation}
where $\theta$ is a set of model parameters, data is the observed data, and \emph{l} is the likelihood. Because the denominator is not a function of $\theta$, we can rewrite Equation \eqref{eq:bayes} as 
\begin{equation}\label{eq:bayes_prop}
	p(\theta|\text{data}) \propto l(\text{data}|\theta) p(\theta).
\end{equation}

Table \ref{tab:BayesCal} shows how each term in Equation \ref{eq:bayes_prop} can be mapped to a component in a calibration exercise. The prior distribution, $p(\theta)$, represents our uncertainty about the distribution of the model parameters before calibrating the model.  Modelers often use various forms of distributions to describe this uncertainty, including beta or logit-normal distribution for probabilities, gamma for rates, or log-normal distributions for rates or hazard ratios. Thus, a prior distribution can be thought of as the uncertainty of the pre-calibrated model input parameters.  For example, a vague distribution can be represented by a uniform distribution where all the values are equally likely within a defined range. Essentially, Bayesian calibration updates this prior distribution based on the observed target data.

The term $p(\theta|data)$ is often referred to as the posterior distribution, which represents the updated distribution of $\theta$ after we observe some data. This is equivalent to the definition of a calibrated parameter when the data are the calibration targets. 

The likelihood function, $l(data|\theta)$, denotes how likely it is that the observed data arise from a given data generation mechanism with a parameter set values $\theta$. From a simulation modeling perspective, $l(data|\theta)$ is equivalent to measuring the goodness of the model output fit to the calibration targets given a simulation model's input parameter set $\theta$. 
Thus, all components of Bayes formula can be mapped to calibration exercises and this formula can be used to obtain the calibrated parameter distributions (a.k.a. the posterior distributions).




While Bayesian calibration provides a seemingly better framework than other calibration approaches, such as through direct-search algorithms, it is rarely used in practice due to computational and practical limitations. In most cases, the computation and technical difficulties preclude applying Bayesian calibration in most realistic models. The main challenge lies in the complexity of applying Equation \ref{eq:bayes_prop}.  Specifically, an analytical solution for $p(\theta|\text{data})$ is unlikely to exist for most realistic simulation models. Thus, specialized algorithms, such as Markov-Chain Monte-Carlo (MCMC) will be necessary which can be both practically challenging and computationally expensive.

\subsection{Metamodels}
To overcome the computational and practical challenges of Bayesian calibration, we propose to use artificial neural network (ANN) metamodels. A metamodel is a second model that approximates the relationship between the simulation model's inputs and outputs (i.e., a metamodel is a model of the model). \cite{Blanning1974, Kleijnen1975a, Kleijnen2005, Kleijnen2015} A metamodel generally describes a simpler relationship among the inputs and outputs than the underlying original simulation model. 
Metamodels range from simple models, such as linear regressions to complex nonlinear models, such as artificial neural networks (ANN).  Although, linear regression models are the most common form of metamodels\cite{Barton2006, Barton2009b,Sacks1989,Fu1994,WeiserFriedman1996,Banks1998,Kleijnen2000,Jalal2013,Jalal2015}, in this paper, we focus on ANN because they are generally more flexible than linear regression and are relatively simple to implement in Stan or BUGS.   

Metamodels are often used because they generally offer vast reduction in computation time.\cite{Kleijnen1979, Friedman1988a,Barton1992,WeiserFriedman1996, OHagan1999, Barton2006, Santos2007, ReisdosSantos2009, Khuri2010} For example, a model that takes several hours or even days to run can be approximated with a metamodel that may only take a few milliseconds to run. This feature has been an attractive attribute of metamodels for many decades in engineering and computer science, however, the use of metamodels has been limited in health decision sciences.\cite{Jalal2013} Some examples of metamodels in health decision sciences have been Gaussian processes to conduct probabilistic sensitivity analysis (PSA) of computationally expensive simulation models, such as microsimulation models.\cite{Stevenson2004, DeCarvalho2019}

While shorter computation time is an important advantage of metamodeling that we will exploit, it is not the only advantage of metamodels in Bayesian calibration. An additional advantage of metamodels is that they can be generalized to models of high complexity, without changing the structure of the metamodel itself. This is because they describe a simpler and more structured relationship than the simulation model. Thus, the same metamodel structure can easily be adapted to more sophisticated simulation models, and the same code in Stan or BUGS can be adapted to more complicated calibration problems with no or minimal change to the code itself.


\subsubsection{ANN metamodels}

\begin{figure}[htbp]
\centering
\fbox{\includegraphics[width=100mm]{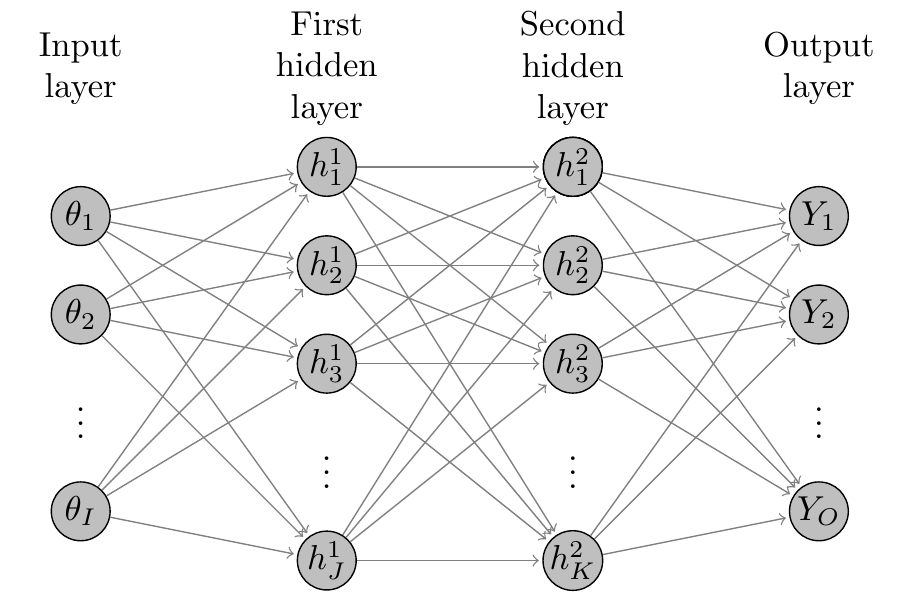}}
\caption{Diagram of general structure of a deep neural network with $I$ inputs, two hidden layers with $J$ and $K$ hidden nodes and $O$ outputs.}
\label{fig:ANN1}
\end{figure}

Artificial neural networks (ANNs) are networks of nonlinear regressions that were originally developed to mimic the neural signal processing in the brain and to model how the nervous system processes complex information \cite{Masson1990,Michie1994,Rojas1996,Jain1996,Olden2008}. In the simplest form, an ANN has three layers of neurons: an input layer, a hidden layer and an output layer. Figure \ref{fig:ANN1} illustrates the basic structure of a four layer neural network with two hidden layers with 4 input nodes in the input layer, 4 nodes in the hidden layer, and 4 output nodes in the output layer. The structure of this ANN can be represented by the following sets of equations   
\begin{equation}\label{eq:ANN}
\begin{split}
	z^{(1)} &= W^{(1)}\theta + b^{(1)} \\
    h &= f^{(1)}\left(z^{(1)}\right) \\
    z^{(2)} &= W^{(2)}h + b^{(2)} \\
    Y &= f^{(2)}\left(z^{(2)}\right), 
\end{split}
\end{equation}
where $\theta$ is the simulation model inputs, $Y$ is the model outputs to be compared to the calibrated targets, and $(W, b) = \left(W^{(1)}, b^{(1)}, W^{(2)}, b^{(2)}\right)$ are the ANN coefficients, such that $W^{(1)}$ are the weights connecting the inputs $\theta$ with the neurons $h$ in the hidden layer and $W^{(2)}$ represents the weights connecting the neurons $h$ in the hidden layer with the output $Y$, $b^{(1)}$ and $b^{(2)}$ are biases connecting the inputs with the neurons in the hidden layer, and the hidden layer with the outputs, respectively, and $f^{(1)}$ is the activation function, which is commonly implemented as sigmoid or logistic functions, such that 
\begin{equation}\label{logistic}
    f^{(1)}\left(z^{(1)}\right)  = \frac{1}{1 + e^{-z^{(1)}}}.
\end{equation}
The function $f^{(2)}$ is often called a transfer function that transforms the results from the neurons of the hidden layer into a working output. The transfer function can also be a sigmoid function or simply a linear function. Thus, the $z^{(1)}$ and $z^{(2)}$ are the weighted sum of inputs from the input layer and the hidden layer, respectively. 

The flexibility of an ANN can be increased by increasing the number of hidden layers and the number of nodes in these layers. ANNs with more than one hidden layer are often referred to as deep ANNs. For this reason, ANNs have recently witnessed significant advances for applications in machine learning, artificial intelligence and pattern recognition.\cite{ravi2016deep}  We use TensorFlow(R) and the package \texttt{keras} in \texttt{R} to create ANN metamodels that approximate the relationship between our model's input parameters and outputs and estimate the coefficients $\beta$ and $W$.\cite{RCoreTeam2018,Jalal2017b} We perform this estimation from a set of probabilistic samples using a Latin hypercube sampling (LHS) design of experiment (DoE) to efficiently sample the input parameter space. Once we obtain the ANN coefficients, we perform the Bayesian calibration using the ANN rather than the simulation model itself.

\subsection{BayCANN algorithm summary}
This section outlines the steps to conduct BayCANN. 

\begin{enumerate}
\item Structure the simulation model such that it produces outputs corresponding to the calibration targets. For example, if calibration data are in the form of disease incidence or prevalence, make sure the model produces these outputs. 
\item Obtain a dataset of parameter sets from a probabilistic sampling framework. This could be done by conducting a Latin hypercube sampling (LHS) of model inputs' prior distributions.
\item Run the simulation model using all the parameter sets from the input samples from the previous step and generate their corresponding simulation model outputs.
\item Train an ANN using a subset of the model inputs and the model outputs, and validate it using the remaining simulation runs. Adjust the ANN's structure to obtain an accurate metamodel.
\item Obtain the calibration targets and, if available, a measure of uncertainty (e.g., standard errors or sample size) from datasets, literature or subjective assessments. 
\item Perform the Bayesian calibration by passing the ANN coefficients, the prior input parameter samples, and the targets data to the ANN framework in Stan. Stan then returns the joint posterior distribution of the calibrated parameters.
\end{enumerate}

The \texttt{R} code provided reflect this algorithm.  In the case study below, we use BayCANN to calibrate a colorectal cancer natural history model.

\subsection{Case study: Natural history model of colorectal cancer}

\begin{figure}[htbp]
\centering
\fbox{\includegraphics[width=\linewidth]{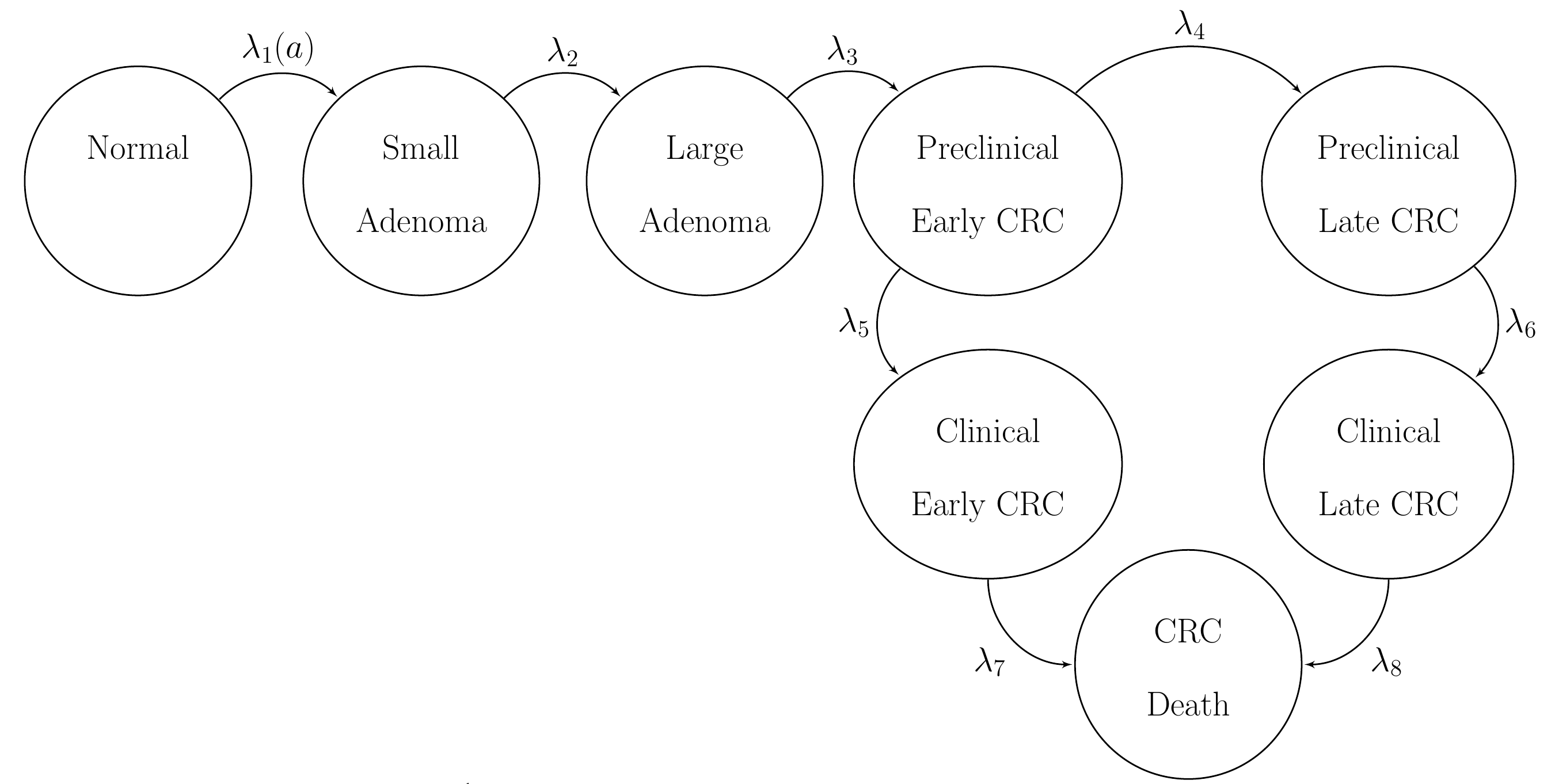}}
\caption{State-transition diagram of the natural history model of colorectal cancer.}
\label{fig:NHM-CRC}
\end{figure}

We use BayCANN to calibrate a state-transition model (STM) of the natural history of colorectal cancer (CRC) implemented in \texttt{R} \cite{Jalal2017b}. We refer to our model as CRCmodR.  CRCmodR is a discrete-time STM based on a model structure originally proposed by Wu et al., 2006 \cite{Wu2006} that has previously been used for testing other methods.\cite{Alarid-Escudero2018a, Heath2020a} Briefly, CRCModR has 9 different health states that include absence of the disease, small  and large precancerous lesions (i.e., adenomatous polyps) , and early and late preclinical and clinical cancer states by stage. Figure \ref{fig:NHM-CRC} shows the state-transition diagram of the model. The progression between health states follows a continuous-time age-dependent Markov process. There are two age-dependent transition intensities (i.e., transition rates), $\lambda_1(a)$ and $\mu(a)$, that govern the age of onset of adenomas and all-cause mortality, respectively. Following Wu's original specification \cite{Wu2006}, we specify $\lambda_1(a)$ as a Weibull hazard such that
\begin{equation}
	\lambda_1(a) = l \gamma a^{\gamma-1},
\end{equation}
where $l$ and $\gamma$ are the scale and shape parameters of the Weibull hazard model, respectively. The model simulates two adenoma categories: small (adenoma smaller than or equal to 1 cm in size) and large (adenoma larger than 1 cm in size). All adenomas start small and can transition to the large size category at a constant annual rate $\lambda_2$. Large adenomas may become preclinical CRC at a constant annual rate $\lambda_3$. Both, small and large adenomas may progress to preclinical CRC, although most will not in an individual’s lifetime. Early preclinical cancers progress to late stages at a constant annual rate $\lambda_4$ and could become symptomatic at a constant annual rate $\lambda_5$. Late preclinical cancer could become symptomatic at a constant annual rate $\lambda_6$. After clinical detection, the model simulates the survival time to death from early and late CRC using time-homogeneous mortality rates, $\lambda_7$ and $\lambda_8$, respectively. In total, the model has nine health states: normal, small adenoma, large adenoma, preclinical early CRC, preclinical late CRC, CRC death and other causes of death. The state-transition diagram of the model is shown in Figure \ref{fig:NHM-CRC}. The model simulates the natural history of CRC of a hypothetical cohort of 50-year-old women in the US over a lifetime. The cohort starts the simulation with a prevalence of adenoma of $p_{adeno}$, from which a proportion, $p_{small}$, correspond to small adenomas and prevalence of preclinical early and late CRC of 0.12 and 0.08, respectively. The simulated cohort in any state is at risk of all-cause mortality $\mu(a)$ which was obtained from the US life tables.\cite{Arias2017}

CRCmodR involves eleven parameters summarized in Table \ref{tab:CRCmodR-input-params}.\cite{Alarid-Escudero2018a}. Mortality rates from early and late stages of CRC ($\lambda_7,\lambda_8]$) could be obtained from cancer population registries (e.g., SEER in the U.S.). Thus, we calibrate the model to the remaining nine parameters ($p_{adeno}$, $p_{small}$,$l$,$\gamma$,$\lambda_2$,$\lambda_3$,$\lambda_4$,$\lambda_5$,$\lambda_6$). 

\begin{figure}[htbp]
\centering
\fbox{\includegraphics[width=\linewidth]{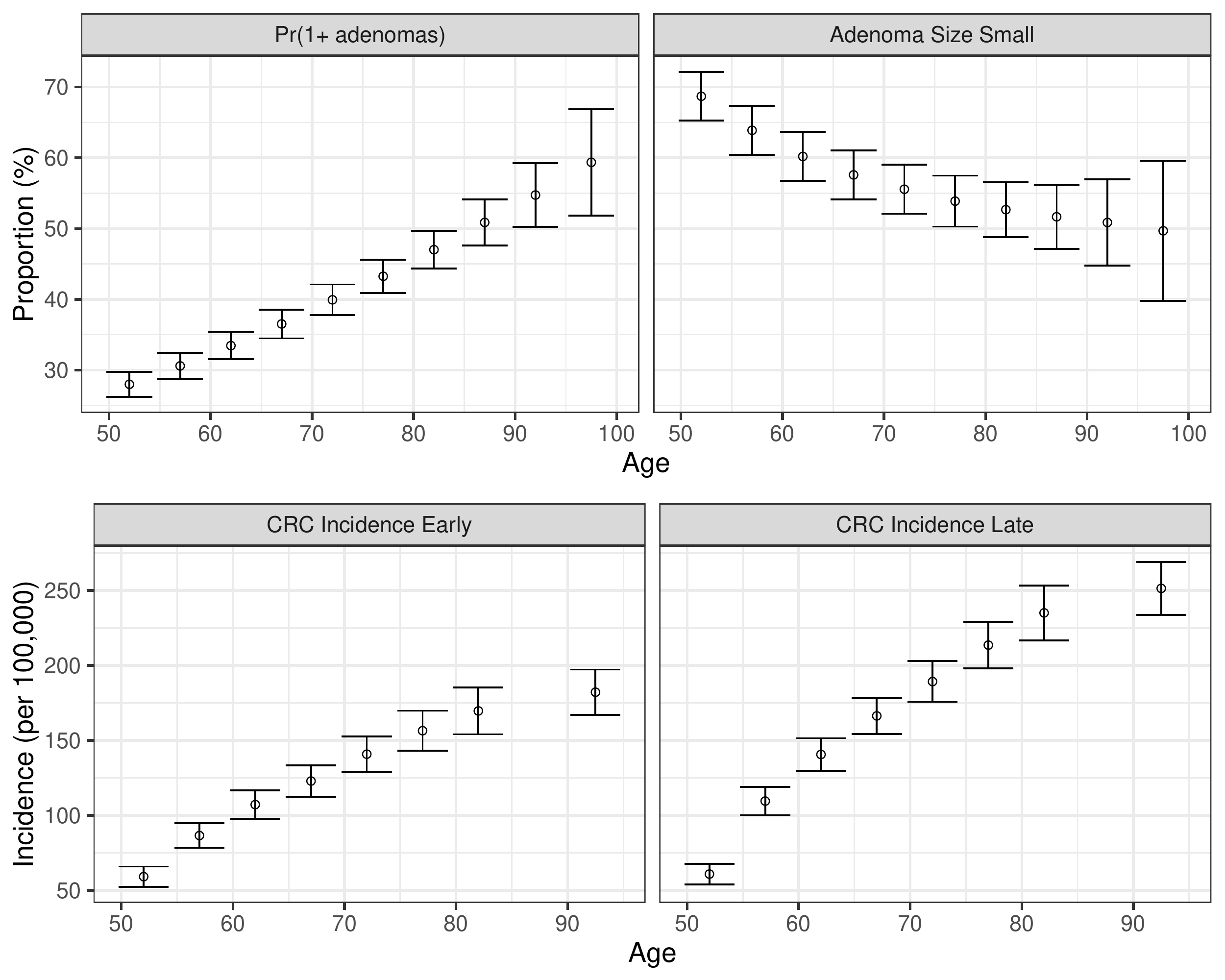}}
\caption{Generated calibration targets and its 95\% CI of a cohort of 500 and 100,000 simulated individuals for adenoma-related targets cancer incidence targets, respectively. These distributions are from 100 different runs using the same parameter set values in each set of runs.}
\label{fig:Calibration-targets}
\end{figure}

To obtain a ``truth'' that we can compare BayCANN against, we first conducted a confirmatory simulation where we \textit{generated} the targets for the base-case values in Table \ref{tab:CRCmodR-input-params}. We generated four different age-specific targets, including adenoma prevalence, proportion of small adenomas and CRC incidence for early and late stages, which resemble commonly used calibration targets for this type of models.\cite{Kuntz2011a} To generate the calibration targets, we ran CRCmodR as a microsimulation \cite{Krijkamp2018} 100 times to produce different adenoma-related and cancer incidence outputs using the base-case values in Table \ref{tab:CRCmodR-input-params}. We then aggregated the results across all 100 outputs to compute their mean and standard errors (SE). Different calibration targets could have different level of uncertainty given the amount of data to compute their summary measures. Therefore, to account for different variations in the amount of data on different calibration targets, we simulated different numbers of individuals for adenoma-related ($N=500$) and cancer incidence ($N=100,000$) individuals. Figure \ref{fig:Calibration-targets} shows the generated adenoma-related and cancer incidence calibration targets aggregated over 100 different runs using the parameter set in Table \ref{tab:CRCmodR-input-params}. 

To create a deep ANN metamodel, we generated a DOE by sampling each of the nine parameters from the ranges of the uniform distributions as shown in Table \ref{tab:CRCmodR-input-params}. We then ran the natural history model and generated model outputs that represent the calibration targets at each of the parameter sets of the DOE.  We ran the model on 10,000 samples from the DOE.  To train the ANN, we divided the DOE dataset into a training subset (8,000 simulations) and a validation subset (2000 simulations) that we used for cross validation.  We define an ANN with two hidden layers and 100 nodes per each hidden layer. Then, we evaluated the performance of the ANN by cross validating the predicted values for the 36 outcomes against the observed values from the validation datasets.  

To calibrate the ANN, we adopted a Bayesian approach that allowed us to obtain a joint posterior distribution that characterizes the uncertainty of both the calibration targets and previous knowledge of the parameters of interest. The likelihood function was constructed by assuming that the targets, $y_{t_i }$, are normally distributed with mean $\phi_{t_i}$ and standard deviation $\sigma_{t_i }$, where $\phi_{t_i} = M[\theta]$ is the model-predicted output for each type of target $t$ and age group $i$ at parameter set $\theta$. 
We defined uniform prior distributions for all $\theta_u$ based on previous knowledge or nature of the parameters (Table \ref{tab:CRCmodR-input-params}). 

To conduct the BayCANN, we implemented the deep ANN in Stan \cite{carpenter2017} which uses a guided MCMC using gradient decent, referred to as Hamiltonian Monte-Carlo.  Similarly, we used the package \texttt{rstan} to conduct the Bayesian calibration in R. In the supplementary material, we provide the R code for wider implementation of our algorithms. In addition, we compare BayCANN against a full Bayesian calibration of the natural history model using the incremental mixture importance sampling (IMIS) algorithm. The IMIS algorithm has been described elsewhere\cite{raftery2010estimating}, but briefly, this algorithm reduces the computational burden of Bayesian calibration by incrementally building a better importance sampling function based on Gaussian mixtures. We compare BayCANN to the IMIS both in terms of accuracy and efficiency.

\section{Results}
We present the performance of the ANN in approximating the output of the simulation model, and compare the joint posterior distribution of the simulation model parameters produced from BayCANN against the IMIS approach.  We compare both the BayCANN and IMIS results recovering the ``truths'' - the parameter values we used to generate the calibration targets.

\subsection{Cross validation}
\begin{figure}[htbp]
\centering
\fbox{\includegraphics[width=\linewidth]{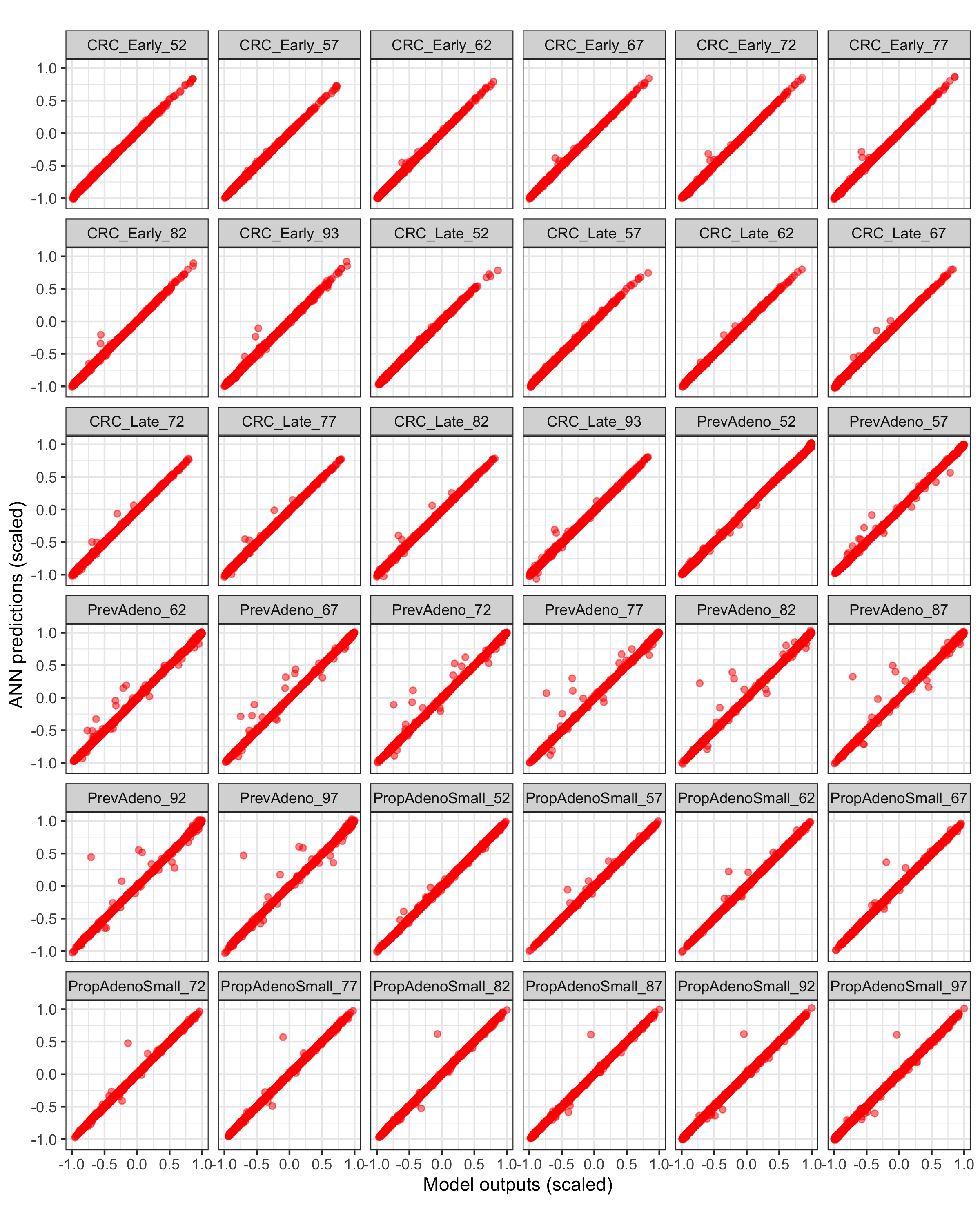}}
\caption{Validation of the fitted ANN on the validation dataset for 36 targets. The x and y axes represent the scaled model outputs and scaled ANN predictions, respectively.}
\label{fig:ANN-validation}
\end{figure}

In the cross-validation exercise, we split the DOE dataset into 80\% training and 20\% validation, to ensure that the ANN can approximate the model outputs generated from inputs that were not used in the training process.  Figure \ref{fig:ANN-validation} illustrates the results of this exercise.  
Each plot represents one of the model outputs, where we compare the ANN's output on the y-axis against the model's output on the x-axis.  
Each red dot represents one of the DOE validation sample that was not used in the training. The ANN had a high prediction performance in approximating the model outputs ($R^2 = 99.9\%$), indicating that the deep ANN is an accurate and reliable metamodel of the simulation model within the parameter ranges.

\subsection{Accuracy and efficiency of BayCANN versus IMIS} 
\begin{figure}[htbp]
\centering
\fbox{\includegraphics[width=\linewidth]{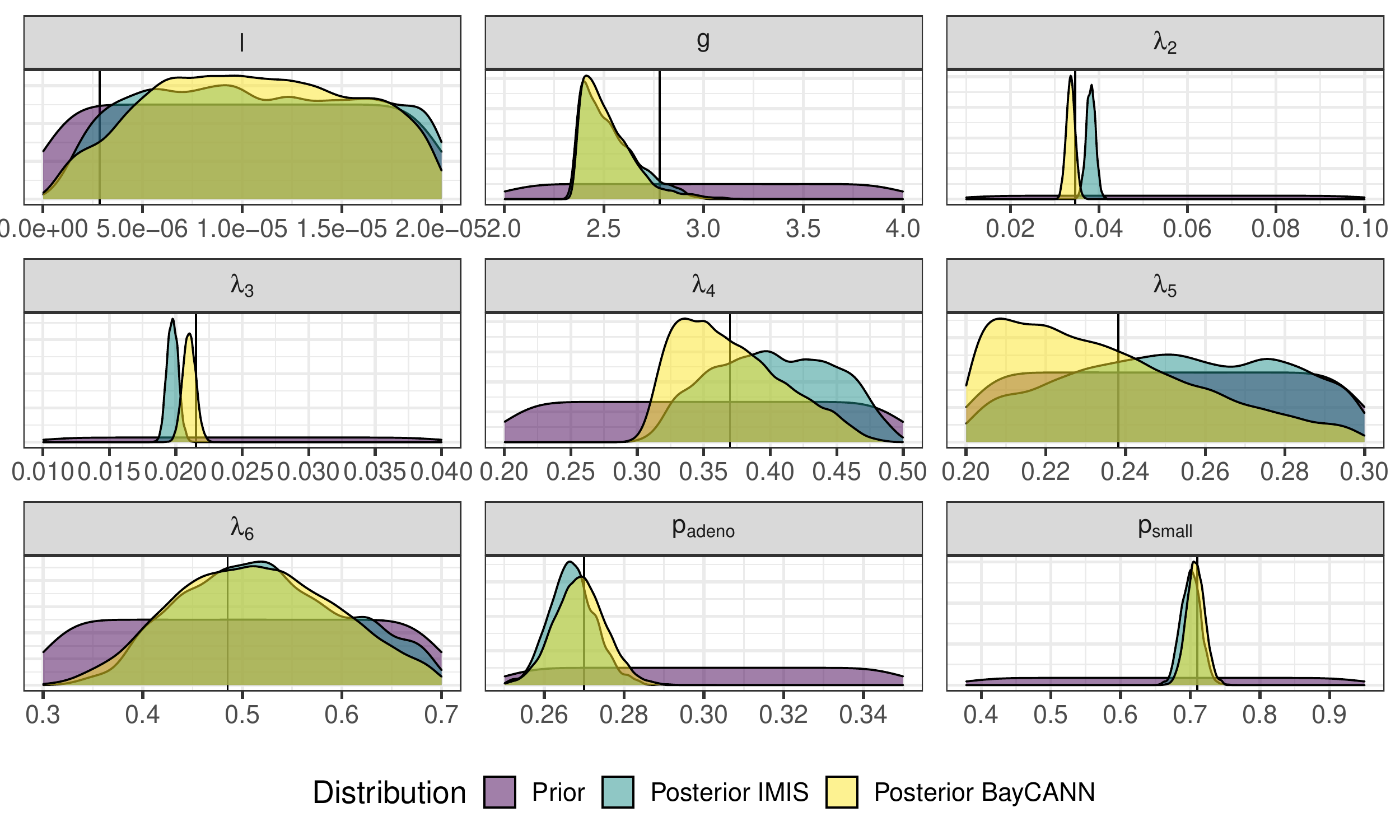}}
\caption{Prior and Marginal posterior distributions of the calibrated parameters from the IMIS and BayCANN methods. The vertical solid lines indicate the value of the parameters used to generate the calibration targets. PDF: probability distribution function; IMIS: incremental mixture importance sampling; BayCANN: Bayesian calibration with artificial neural network metamodeling.}
\label{fig:Marginal-post}
\end{figure}

Figure \ref{fig:Marginal-post} compares BayCANN against IMIS in recovering the true parameter values used to generate the targets.  The posterior distributions for both methods overlap and cover the truth for all parameters in the case of the BayCANN method and most of the parameters in the case of the IMIS.  Table \ref{tab:compare_ANN_IMIS} compares the ANN against the IMIS method for recovering the true parameter values.  This table presents the mean of the calibrated parameters and the absolute deviation from the truth.  The ratio of the ANN deviation to the IMIS deviation presents the relative performance of the two methods.  This ratio of deviations is less than 1 for most parameters except for $g$ which is slightly above one, indicating that the mean of the ANN distribution performed better at recovering the true parameter value. For the majority of the parameters (six out of nine) this ratio was less than 0.35, indicating a better accuracy for BayCANN relative to IMIS.  

In addition, BayCANN was five times faster than the IMIS. The IMIS algorithm 
took 80 minutes to run in a MacBook Pro Retina, 15-inch, Late 2013 with a 2.6 GHz Intel Core i7 processor with 4 cores and 16 megabytes of RAM.  The Bayesian ANN took only 15 minutes on the same computer; 5 minutes to produce 10,000 samples for the DOE dataset and 10 minutes to fit the ANN and produce the joint posterior distributions.  


\section{Discussion}
In this study, we propose BayCANN as a feasible and practical solution to the challenges of Bayesian calibration in health decision science models. We compared the accuracy and efficiency of BayCANN approach against the IMIS algorithm using a natural history model of colorectal cancer.  In our case study, BayCANN was both faster and overall more accurate in recovering the true parameter values than the IMIS algorithm. We developed BayCANN approach to be generalizable to models of various complexities, and we provide the open source implementation in R and Stan to facilitate its wider adoption.

Bayesian calibration is superior to other forms of calibration (such as direct-search algorithms) because it can reveal the joint posterior distribution of the calibrated parameters.\cite{rutter2019microsimulation} 
This joint distribution will be informative in the case of non-identifiablity where calibration targets are not sufficient to provide a unique solution to the calibrated parameters.\cite{Alarid-Escudero2018a} Non-identifiability is often overlooked using standard non-Bayesian calibration approaches.  In addition, Bayesian calibration provides other practical advantages because the samples from the joint posterior distribution can be used directly as inputs to probabilistic sensitivity analyses (PSA) of cost-effectiveness models and other models.  

Despite its advantages, Bayesian calibration has rarely been used in calibrating models in health decision sciences because of practical and computational burdens of its implementation.  The practical burdens involves adapting the simulation model to the probabilistic programming languages or other algorithms that are often designed for simple regression-type models. The computational burdens involves, the time it might take for algorithms such as Markov-chain Monte-Carlo (MCMC) to converge especially given the complexity of simulation models.  These two factors are the main reasons for the limited adoption of Bayesian calibration in practice. 

We illustrated how ANNs can be used as a practical solution to these computational and technical challenges.  
In this application we use an ANN to ``learn'' the relationship between the model inputs and outputs using a training subset of the simulation runs.  We used cross-validation to avoid over-fitting the ANN to the calibration targets.\cite{Kilmer1994a,Syberfeldt2008,Khadra2013}  
The ANN was an efficient and accurate emulator of the natural history model because it predicted the outputs of our simulation model with high accuracy and in a fraction of the time the original simulation model required. Compared to linear regressions metamodels, ANN metamodels are more powerful, require fewer assumptions and less precise information about the system to be modeled (i.e., the degree of polynomials).\cite{Alam2004,Padgett1992}  These properties make it an attractive choice for Bayesian calibration.

In addition to the computational and practical advantages of using ANNs, BayCANN may have an additional advantage for representing models with first-order Monte-Carlo noise from individual-based state-transition models (iSTM). Traditionally, calibrating these models has been especially challenging because of (1) the stochasticity of each simulation due to the output of the simulation varying given the same set of input parameter values, and (2) the extra computational burden involved in calibrating iSTM. Because BayCANN averages over a set of simulations, it can account for the first order Monte-Carlo noise.  

Our approach has some limitations. First, accuracy - Because ANN's are metamodels, they may rarely achieve 100\% precision compared to using the simulation model itself.  In our example, with a relatively simple ANN (only two hidden layers with 100 hidden nodes each), we were able to achieve 99.9\% accuracy.  However, for other application, the accuracy of the ANN might be lower.  In addition, over-fitting can be a serious problem with any metamodel especially when the purpose of the metamodel is as sensitive as calibration.  To reduce the chance of overfitting, we cross-validated the model against a subset of the data, which is a commonly used technique.  We used 80\% of the 10,000 PSA observations for training the ANN, and 20\% to validate the ANN.  We visually inspected the degree of fit for the simulation output against those predicted by the ANN (Figure \ref{fig:ANN-validation}). Second, similar to any Bayesian model, the choice of priors could be important. Fortunately, in simulation models, modelers often make careful choices of their priors when they design their models and run PSA analyses. Thus, depending on the ranges chosen, the best-fitting parameters may be outside the simulated ranges. Importantly, the joint posterior distribution can give insights into the parameter ranges.  For example, if a parameter is skewed heavily without a clear peak, that may indicate that the parameter range needs to be shifted to cover values that may fit better.  This process is usually iterative and may involve multiple steps or redefining the parameter ranges and recalibrating the model. Finally, there is no strict guideline for choosing the number of hidden ANN layers or the number of nodes per layer. In this study, we chose an ANN with two hidden layers and 100 nodes per layer.  Adjusting these parameters and additional parameters of the Bayesian calibration process can provide better calibration results.  While determining these values apriori can be challenging, we recommend modelers who wish to use BayCANN to start with simple settings initially and gradually increase the complexity of the ANN to accommodate their particular needs.  We provide flexible code in R and Stan to simplify these tasks.

In summary, Bayesian calibration can reveal important insights into model parameter values and produce outcomes that match observed data.  BayCANN is one effort to target the computational and technical challenges of Bayesian calibration for complex models.  

\section{Acknowledgements}
Dr. Jalal was supported by the Center for Disease Control and Prevention Contract No. 34150, and a grant from the National Institute on Drug Abuse of the National Institute of Health under award no. K01DA048985. Dr Alarid-Escudero was supported by a grant from the National Cancer Institute (U01-CA-253913-01) as part of the Cancer Intervention and Surveillance Modeling Network (CISNET), the Gordon and Betty Moore Foundation, and Open Society Foundations (OSF).

\bibliographystyle{plain}
\bibliography{main.bib}

\newpage

\section{Tables}
\begin{table}[htbp]
\centering
\caption{Contrasting Bayes formula with a calibration process.}
\footnotesize
\begin{tabular}{lL{3in}L{3in}}
\toprule
	Term  & Bayesian Context & Calibration Context \\ \hline
\midrule
    $p(\theta)$  & Prior distribution of the model input parameters $\theta$ & Pre-calibrated model input parameters \\ \hline
	$p(\theta|\text{data})$  & Posterior distribution of the model parameters $\theta$ given observed data & Calibrated model parameters to target data \\ \hline
	$l(\text{data}|\theta)$  & Likelihood of the data given model parameters $\theta$ & Goodness-of-fit measure; how well the model output fits the target data given a particular value of $\theta$ \\ \hline
\bottomrule
\end{tabular}
  \label{tab:BayesCal}
\end{table}

\newpage

\begin{table}[H]
\footnotesize
\caption{\label{tab:CRCmodR-input-params} The parameters of the natural history
model of colorectal cancer (CRC).  The base valeus are used to generate the calibration targets and the ranges of the uniform distribution used as priors for the Bayesian calibration.}

\begin{tabular}{llcccc}
\toprule
\textbf{Parameter} & \textbf{Description} & \textbf{Base value} &
\textbf{Calibrate?} & \textbf{Source} & \textbf{Prior range} \\
\midrule
\(l\) & Scale parameter of Weibull hazard & 2.86e-06 & Yes &
\cite{Wu2006} & $[2\times10^{-6},2\times10^{-5}]$  \\ 
\(g\) & Shape parameter of Weibull hazard & 2.78 & Yes &
\cite{Wu2006} & $[2.00,4.00]$ \\ 
\(\lambda_2\) & Small adenoma to large adenoma & 0.0346 & Yes &
\cite{Wu2006} & $[0.01,0.10]$ \\ 
\(\lambda_3\) & Large adenoma to preclinical early CRC & 0.0215 & Yes &
\cite{Wu2006} & $[0.01,0.04]$ \\ 
\(\lambda_4\) & Preclinical early to preclinical late CRC & 0.3697 & Yes & 
\cite{Wu2006} & $[0.20,0.50]$ \\ 
\(\lambda_5\) & Preclinical early to clinical early CRC & 0.2382 & Yes &
\cite{Wu2006} & $[0.20,0.30]$ \\ 
\(\lambda_6\) & Preclinical late to clinical late CRC & 0.4852 & Yes &
\cite{Wu2006} & $[0.30,0.70]$ \\ 
\(\lambda_7\) & CRC mortality in early stage & 0.0302 & No &
\cite{Wu2006} & - \\ 
\(\lambda_8\) & CRC mortality in late stage & 0.2099 & No &
\cite{Wu2006} & - \\ 
\(p_{adeno}\) & Prevalence of adenoma at age 50 & 0.27 & Yes &
\cite{Rutter2007} & $[0.25,0.35]$ \\ 
\(p_{small}\) & Proportion of small adenomas at age 50 & 0.71 & Yes &
\cite{Wu2006} & $[0.38,0.95]$ \\ 
\bottomrule
\end{tabular}
\end{table}
\newpage


\begin{table}
\scriptsize
\caption{\label{tab:compare_ANN_IMIS} Comparing the accuracy of BayCANN against IMIS in recovering the true parameter values as shown in Table \ref{tab:CRCmodR-input-params}} 

\begin{tabular}{@{}lllllll@{}}
\toprule
\textbf{Parameter} & \textbf{Truth} & \textbf{ANN (mean)} & \textbf{IMIS (mean)} & \textbf{ANN (deviation)} & \textbf{IMIS (deviation)} & \textbf{Deviation ratio} \\ 
\midrule
\(\l\)          & 0.0000029      & 0.0000107           & 0.0000108            & 0.0000078                & 0.0000079                 & 0.9861448                \\
g                  & 2.7790840      & 2.5249560           & 2.5293840            & 0.2541280                & 0.2497000                 & 1.0177333                \\
\(\lambda_2\)          & 0.0346000      & 0.0336306           & 0.0381652            & 0.0009694                & 0.0035652                 & 0.2719139                \\
\(\lambda_3\)          & 0.0215000      & 0.0210140           & 0.0197588            & 0.0004860                & 0.0017413                 & 0.2791156                \\
\(\lambda_4\)          & 0.3697000      & 0.3699218           & 0.4026561            & 0.0002218                & 0.0329561                 & 0.0067302                \\
\(\lambda_5\)          & 0.2382000      & 0.2343703           & 0.2528302            & 0.0038297                & 0.0146302                 & 0.2617668                \\
\(\lambda_6\)          & 0.4852000      & 0.5137273           & 0.5231138            & 0.0285273                & 0.0379138                 & 0.7524252                \\
\(p_{adeno}\)         & 0.2700000      & 0.2689944           & 0.2669941            & 0.0010056                & 0.0030059                 & 0.3345421                \\
\(p_{small}\)     & 0.7100000      & 0.7078401           & 0.7008754            & 0.0021599                & 0.0091246                 & 0.2367117                \\ 
\bottomrule
\end{tabular}
\end{table}

\clearpage

\section*{Figures}

\newpage 

\clearpage

\begin{figure}[htbp]
\centering
\fbox{\includegraphics[width=\linewidth]{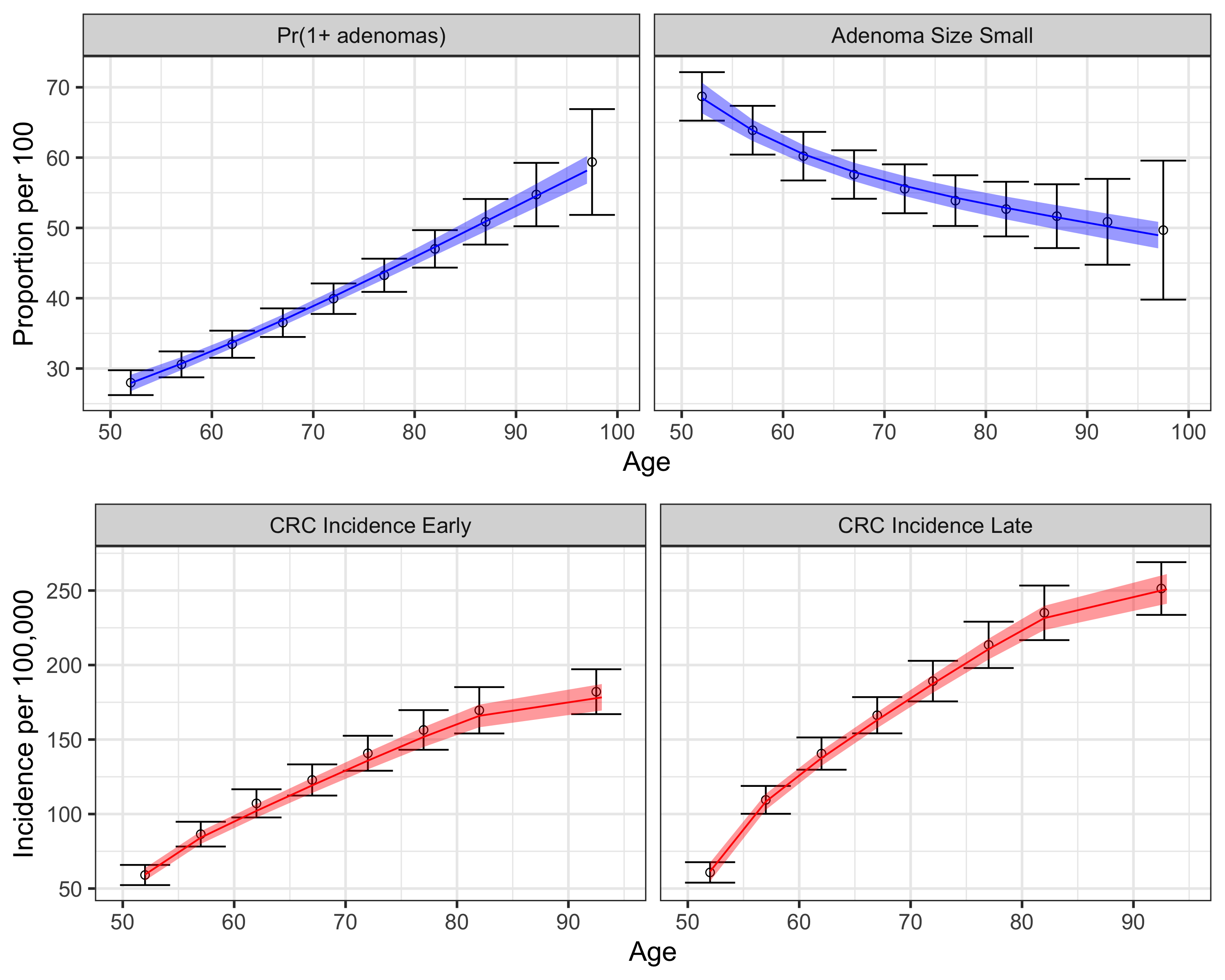}}
\caption{BayCANN calibration results. The upper panel show  adenoma targets and lower panels show cancer incidence targets by stage. Calibration targets with their 95\% confidence intervals are shown in black. The colored curves show the posterior model-predicted mean, and the shaded area shows the corresponding 95\% posterior model-predictive credible interval of the outcomes. }
\label{fig:posterior-predicted-output}
\end{figure}


\end{document}